\documentclass[pra,twocolumn,groupedaddress,showpacs,floatfix]{revtex4}
\usepackage{graphics}
\usepackage{graphicx}
\usepackage{bm}
\usepackage{amsmath}
\usepackage{amsfonts}
\usepackage{amssymb}
\usepackage{latexsym}
\usepackage{color}
\usepackage{bbold}
\usepackage{braket}

\begin{document}

\title{Cavity-mediated collective laser-cooling of a non-interacting atomic gas \\ inside an asymmetric trap to very low temperatures}
\author{Oleg Kim,$^1$ Prasenjit Deb$\,^{1,2}$ and Almut Beige$\,^1$}
\address{$^1$The School of Physics and Astronomy, University of Leeds, Leeds LS2 9JT, United Kingdom \\
$^2$Department of Physics and Centre for Astroparticle Physics and Space Science, Bose Institute, Bidhan Nagar Kolkata 700091, India}
\date{\today}

\begin{abstract}
In this paper, we identify a many-particle phonon expectation value $\zeta$ with the ability to induce collective dynamics in a non-interacting atomic gas inside an optical cavity. We then propose to utilise this expectation value to enhance the laser cooling of many atoms through a cyclic two-stage process which consists of cooling and displacement stages. During cooling stages, short laser pulses are applied. These use $\zeta$ as a resource and decrease the vibrational energy of the atomic gas by a fixed amount. Subsequent displacement stages use the asymmetry of the trapping potential to replenish the many-particle phonon expectation value $\zeta$. Alternating both stages of the cooling process is shown to transfer the atomic gas to a final temperature which vanishes in the infinitely-many particle limit.
\end{abstract}

\maketitle

\section{Introduction} \label{introduction}

Laser sideband cooling allows for cooling single, strongly-confined atomic particles to very low temperatures \cite{WinelandDehmelt,Norah}. Its discovery opened the way for experiments which test the foundations of the quantum physics of single particles and have applications ranging from quantum metrology to quantum computing \cite{RevMod}. However, despite imminent quantum technological applications of ultra-cold many-body systems, applying laser cooling to large ensembles of particles or more complex species, like molecules, is not straightforward. For example, the most efficient technique to cool atomic gases to the low temperatures required for Bose Einstein condensation is evaporative cooling, which involves the removal of a significant percentage of particles from the trap. Alternative cooling techniques with the potential of maintaining the initial number of trapped particles therefore receive a lot of attention in the literature \cite{Lev}. 

For example, it is hoped that cavity-mediated laser-cooling will enable us to cool large samples of atomic particles collectively and non-destructively to very low temperatures. First indications that this might indeed be the case were found in Paris already more than twenty years ago \cite{Karine}. More systematic experimental studies of cavity-mediated laser cooling have subsequently been reported by several groups \cite{pinkse2,Rempe,vuletic,new,kimble,chap,Meschede,Wolke,Barrett}. All of these experiments have their respective merits. However, all of them are essentially equivalent to laser-cooling without optical cavities and do not provide a much more efficient cooling mechanism for many particles. 

The theory of cavity-mediated laser cooling of free particles was first discussed by Mossberg {\em et al.}~\cite{lewen} and Zaugg {\em et al.}~\cite{lewen2}. Later, Ritsch and collaborators \cite{peter2,peter3} and others \cite{Murr,vuletic10,Robb} developed semiclassical theories to model cavity-mediated cooling processes. In 1993, Cirac {\em et al.}~\cite{Cirac2} introduced a master equation approach to analyse cavity-mediated laser-cooling in more detail. Since the precision of calculations, which are based on master equations, is easier to control than the precision of semiclassical approximations, this approach has subsequently been used by many authors to emphasize a close analogy between laser-sideband and cavity-mediated laser-cooling of atomic particles \cite{Cirac4,cool,morigi,Morigi,Tony3,kim1}. 

Recently, Mishina showed that the cooling dynamics of particles can be affected by trap inhomogeneities \cite{Mishina}. In this paper, we adopt an alternative but conceptually related approach to cool a non-interacting atomic gas inside an optical cavity. Instead of considering a symmetric harmonic trap, we place the particles into an {\em asymmetric} trapping potential. The cooling process that we propose here is qualitatively different from other laser cooling techniques \cite{Cirac2,Cirac4,cool,morigi,Morigi,Tony3,kim1} for which the final energy of the atoms is usually independent of the number $N$ of particles inside the trap. In contrast to this, we predict a final mean phonon number which vanishes in the infinitely-many particle limit. Hence the proposed cooling mechanism has the potential to transfer an atomic gas to very low temperatures. Notice that the paper by Mishina \cite{Mishina} predicts a similarly strong but less useful dependence of the final temperature of the atoms on the particle number. 

To illustrate our main ideas we first consider an idealised version of the proposed collective cooling mechanism. Our initial toy model is relatively simple and cannot be implemented physically but shows clearly what we aim to achieve, at least approximately, with the experimental setup which we analyse later on in this paper. Suppose $N$ non-interacting atomic particles are so strongly confined that their motion becomes quantised. Moreover we assume in the following that the atomic gas is placed inside an optical cavity and that its system Hamiltonian equals
\begin{eqnarray} \label{Heff}
H_{\rm ideal} &=& \sum_{i=1}^N \hbar J \left[ b_i c^\dagger + b_i^\dagger c \right] 
\end{eqnarray}
with respect to an accordingly chosen interaction picture. Here $b_i$ with $[ b_i,b_j^\dagger ] = \delta_{i,j}$ and $c$ with $[ c,c^\dagger] = 1$ are the bosonic annihilation operators of a single phonon in the motion of particle $i$ and of a photon inside the resonator, respectively, and $J$ denotes a phonon-photon coupling rate. Moreover, we assume that the field inside the cavity has a finite spontaneous decay rate $\kappa$. Consequently, the density matrix $\rho_{\rm I}$ of phonons and cavity photons evolves in the corresponding interaction picture according to a Markovian master equation in Lindblad form, 
\begin{eqnarray} \label{1}
\dot{\rho}_{\rm I} &=& - {{\rm i} \over \hbar} \left[ H_{\rm ideal} , \rho_{\rm I} \right] + {\kappa \over 2} \left( 2 c \, \rho_{\rm I} \, c^\dagger - c^\dagger c \, \rho_{\rm I} - \rho_{\rm I} \, c^\dagger c \right) \, . ~~~~
\end{eqnarray}
This master equation can be used to predict the dynamics of expectation values. In the following, we are especially interested in the two expectation values
\begin{eqnarray} \label{notation}
m &\equiv & {1 \over N} \sum_{i=1}^N \langle b_i^\dagger b_i \rangle \, , \nonumber \\
\zeta &\equiv & {1 \over N(N-1)} \sum_{i=1}^N  \sum_{j \neq i} \langle b_i^\dagger b_j \rangle \, . ~
\end{eqnarray}
Both variables are normalised such that they are independent of $N$ in the infinitely many-particle limit. Proceeding as described in App.~\ref{App}, one can then show that 
\begin{eqnarray} \label{rates}
\dot{m} = \dot{\zeta} &=& - {4 N J^2 \over \kappa} \, \zeta 
\end{eqnarray}
in case of a relatively large cavity decay rate $\kappa$ and in case of a sufficiently large $N$, 
\begin{eqnarray} \label{5}
\kappa \gg J ~~ {\rm and} ~~ N \gg 1 \, .
\end{eqnarray}
This implies cooling, if $m$ is at least an approximate measure for the vibrational energy of the trapped particles. The derivation of the above effective rate equations only involves standard approximations which are well justified in the parameter regime given in Eq.~(\ref{5}). In this regime, the term on the right hand side of Eq.~(\ref{rates}) dominates the time evolution of $m$ and $\zeta$ as long as $\zeta$ is non-zero and sufficiently large.

\begin{figure}[t]
\center
\includegraphics[width=80mm]{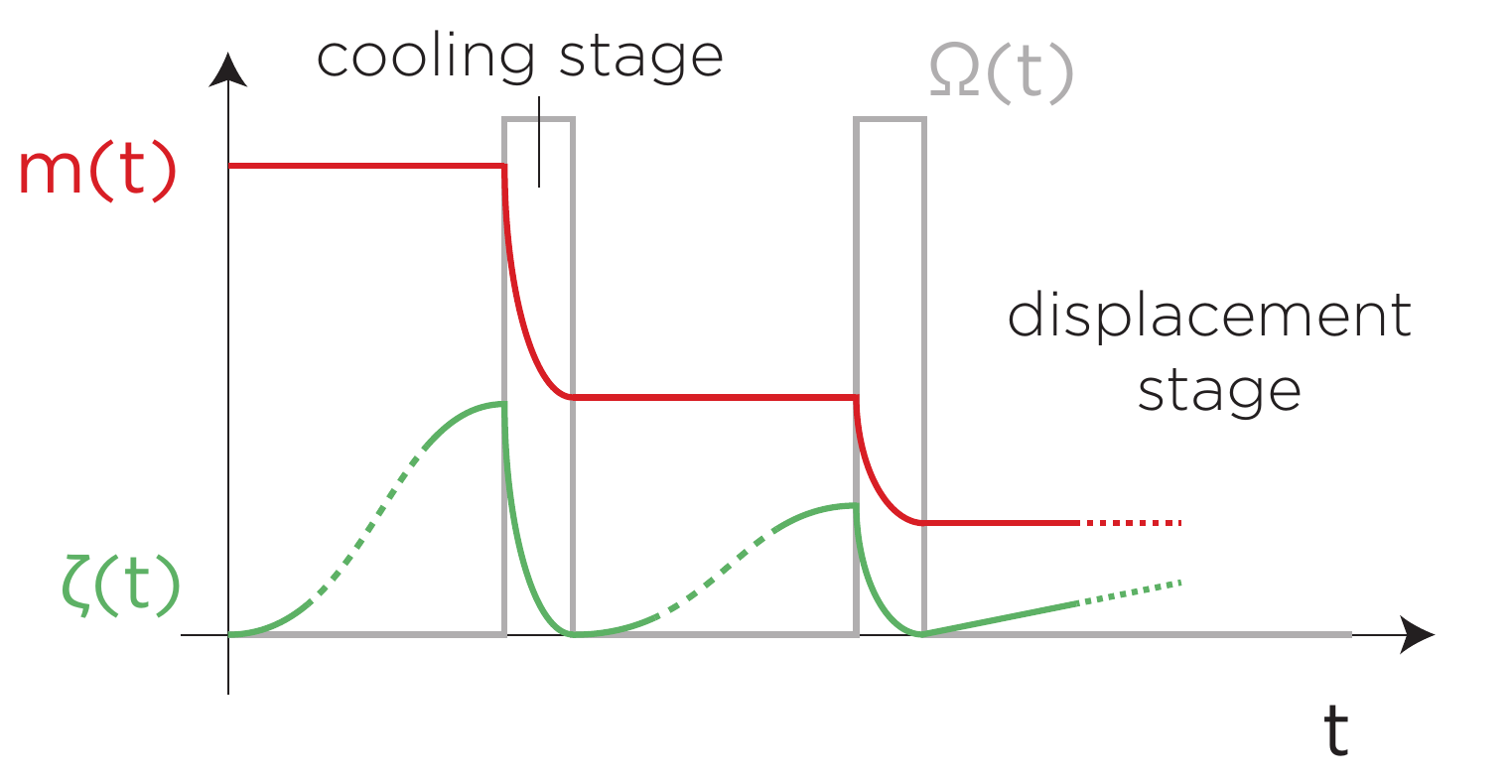} \\[-0.5cm]
\caption{Schematic view of the dynamics of the mean phonon number $m$ and the mean phonon expectation value $\zeta$, which is a measure for the average distance of the atoms from the centre of the trapping potential. The proposed cooling process consists of cooling stages interspersed with displacement stages. During each displacement stage, the atoms move freely inside an asymmetric trap and accumulate some distance away from the centre of the trap, thereby accumulating a non-zero value for $\zeta$, while $m$ remains essentially constant. During each cooling stage, a cooling laser with Rabi frequency $\Omega$ is turned on which returns the atoms to the centre of the trap, thereby returning $\zeta$ to zero while reducing $m$ with a collectively-enhanced cooling rate.} \label{timeline}
\end{figure}

Notice also that the above-described collective dynamics is the result of the presence of many particles. It is qualitatively different from the dynamics of the individual system components. For example, for $N=1$, Eq.~(\ref{notation}) automatically implies $\zeta \equiv 0$. Hence, for a single particle, we always have $\dot m = \dot \zeta = 0$ on the fast time scale. However, for $N \gg 1$ and when $\zeta$ is sufficiently large, the above-described collective dynamics dominates all other dynamics and the mean phonon number $m$ changes relatively rapidly. Suppose $\zeta (0)$ is positive at $t=0$, when the ideal Hamiltonian $H_{\rm ideal}$ is switched on. Solving Eq.~(\ref{rates}), we then find that  
\begin{eqnarray} \label{msmall}
m(t) &=& m(0) - \zeta (0)  \, , \nonumber \\
\zeta (t) &=& 0
\end{eqnarray}
after a relatively short time $t \gg \kappa / 4 N J^2$. Using this dynamics, it is in principle possible to remove all vibrational energy from the atomic particles and to cool an atomic gas to very low temperatures. To achieve this, the quantum resource $\zeta$, which gets used up in the cooling process, needs to be continuously replenished. This can be done through a cyclic two-step process which consists of cooling and displacement stages, as illustrated in Fig.~\ref{timeline}. The cooling stages employ the variable $\zeta$ to rapidly decrease the vibrational energy of the atomic gas. The purpose of the displacement stages is to restore a non-zero and positive phonon expectation value $\zeta$.

Next, let us discuss in more detail, how to replenish $\zeta$, whenever this resource has been used up, without increasing the vibrational energy of the trapped particles. As we shall see below, $\zeta$ is essentially a measure for the mean distance of the atoms from the centre of the trap. One way of solving this task is to employ an asymmetric trapping potential and to let the atoms evolve freely between cooling stages. This is exactly what we propose here. During displacement stages, the atoms oscillate freely within the asymmetric trap and naturally accumulate some distance away from its centre. There is nothing quantum about their motion. The hotter the atoms are, the further they move, while their vibrational energy remains the same.

\begin{figure}[t]
\center
\includegraphics[width=80mm]{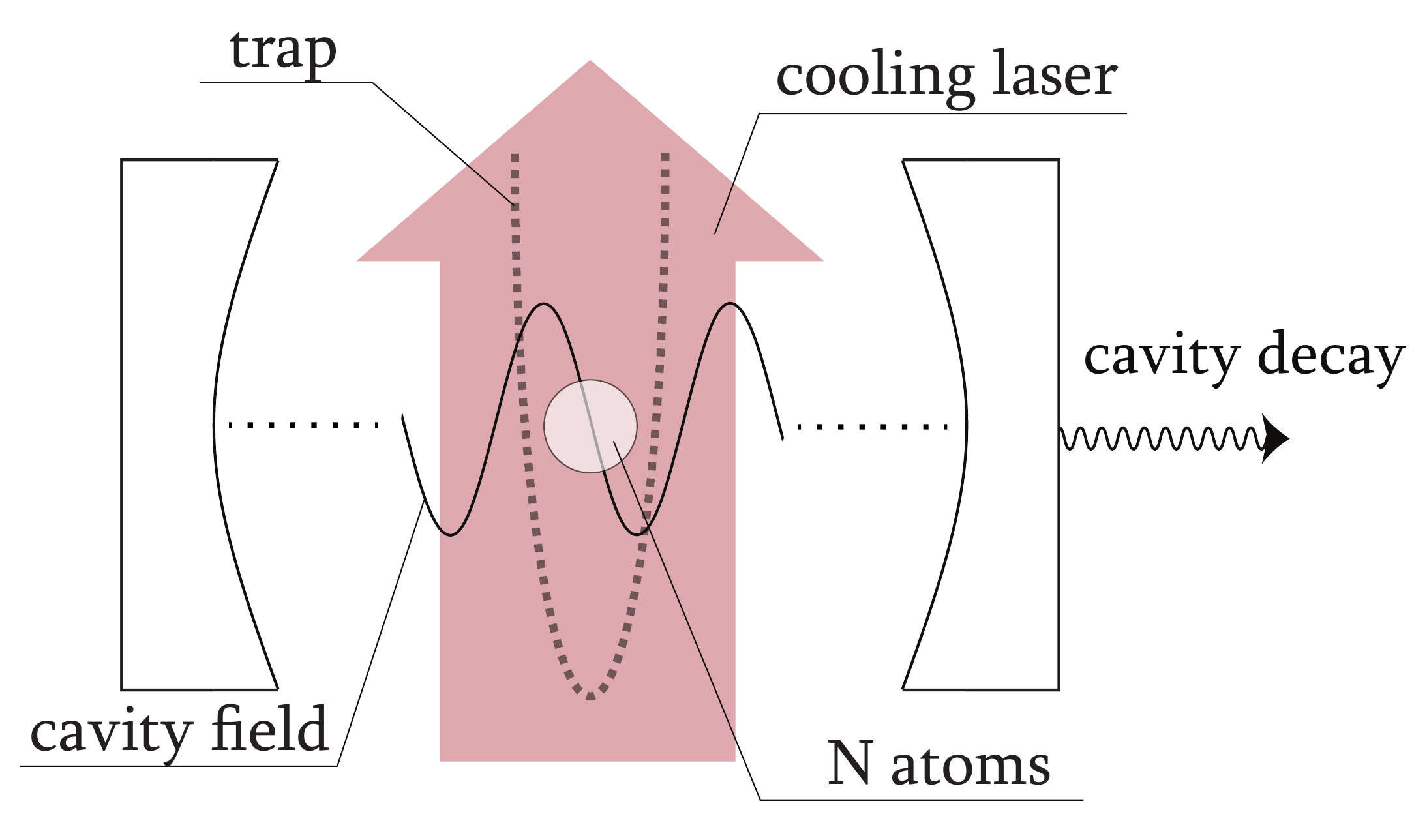} 
\caption{Schematic view of the proposed experimental setup. This consists of an atomic gas which is externally confined such that the centre of the trap coincides with the node of an optical cavity. The motion should become quantised in the direction of the cavity axis. The trapping potential is assumed to be only approximately harmonic. It should exhibit an asymmetry, which places freely-moving atoms on average slightly away from the centre of the trap. To cool the atomic gas, a series of short laser pulses is applied, which enter the cavity from the side.} \label{setup}
\end{figure}

\begin{figure}[t]
\center
\includegraphics[width=80mm]{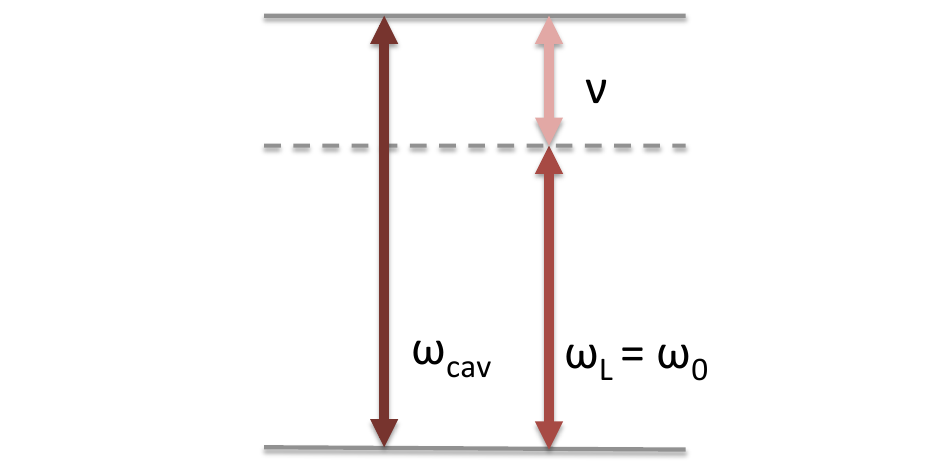} 
\caption{Schematic view of the relevant frequencies of the experimental setup in Fig.~\ref{setup}. To maximise the cooling process, the energy of the cavity photons $\hbar \omega_{\rm cav}$ should equal the sum of the energy of a single phonon $\hbar \nu$ and the energy of an atomic excitation $\hbar \omega_0$. In addition we assume resonant laser driving of the atoms with frequency $\omega_{\rm L}$ such that $\omega_{\rm L} = \omega_0$.} \label{Fig3}
\end{figure}

Moreover, we need to find a way to realise the system Hamiltonian $H_{\rm ideal}$ in Eq.~(\ref{Heff}), at least to a very good approximation. To do so, we proceed as usual in cavity-mediated laser cooling \cite{Cirac4,cool,morigi,Morigi,Tony3,kim1}. During each cooling stage, a cooling laser with Rabi frequency $\Omega$ is applied. The cooling laser should enter the resonator from the side, as illustrated in Fig.~\ref{setup}. Moreover, the laser frequency $\omega_{\rm L}$ should coincide with the atomic transition frequency $\omega_0$ but should be below the cavity photon frequency $\omega_{\rm cav}$. As we shall see below, best results are obtained when the cavity-atom detuning equal the phonon frequency $\nu$ of the trapped particles, 
\begin{eqnarray} \label{delta}
\omega_{\rm cav} - \omega_0 &=& \nu \, ,
\end{eqnarray}
as illustrated in Fig.~\ref{Fig3}. Given this condition, the creation of a cavity photon is most likely accompanied by the loss of a phonon and vice versa, while the atoms remain essentially in a stationary state. Important for the proposed cooling scheme to work is that the atoms are strongly confined. They should be trapped well within an optical wavelength (see e.g.~Refs.~\cite{Rempe2,Rempe3,Rempe4}) and the centre of their asymmetric trap should coincide with a node of the cavity field. Moreover, the trap asymmetry should be relatively weak such that the trapping potential is approximately harmonic. 

For most cavity-mediated laser cooling schemes, it is the ratio between cooling and heating rates which determines the final temperature of the atomic gas (see e.g.~Ref.~\cite{kim1}). Any attempt to collectively increase cooling rates usually also results in a collective enhancement of heating rates. As a result, the cooling process might become faster but the final vibrational energy per particle usually remains the same. In contrast to this, for the cooling scheme which we propose here, the time derivative of the expectation value $m$ is always negative. Only cooling processes are collectively enhanced. Heating processes become secondary and the final temperature of the particles is no longer restricted by the system parameters. Alternating displacement and cooling stages, as illustrated in Fig.~\ref{timeline}, should therefore allows us to transfer the atomic gas to very low temperatures. Cooling only stops when the particles are so cold that they no longer change their position during displacement stages. As we shall see below, the final mean phonon number $m_{\rm final}$ per particle scales as $1/N^{1/2}$ and tends to zero in the infinitely many-particle limit.

There are five sections in this paper. Section \ref{theory} discribes the proposed experimental setup and introduces its theoretical model. In Section \ref{analysis}, we analyse the dynamics of the proposed cyclic two-stage cooling process in detail. The asymmetry of the trap and other effects, like detunings and additional interaction terms, which are not present in the above simplified toy-model  are explicitly taken into account. We derive effective rate equations for the time evolution of the mean phonon expectation values $m$ and $\zeta$ during cooling and displacement stages. In Section \ref{estimate}, we estimate the general scaling of the final mean phonon number $m_{\rm final}$ of the proposed cooling scheme. Finally we summarise our findings in Section \ref{conclusions}. Some calculations are confined to appendices.

\section{Experimental setup and theoretical models} \label{theory}

In this section, we discuss the experimental setup in Fig.~\ref{setup} in more detail and introduce the theoretical models which allows for analysing the dynamics of the atomic gas throughout the proposed cooling process. 

\subsection{Displacement stage} \label{displ}

As mentioned already above, the cooling laser is turned off during each displacement stage. When this happens, there is no interaction between vibrational and other degrees of freedom of the atoms. Only the Hamiltonian describing the motion of the particles has to be taken into account when analysing the dynamics of phonon expectation values during this stage. In the following, we assume that the external trap confines the atoms in the direction of the cavity axis. In this case, the Hamiltonian for the vibrational energy of the atomic gas can be written as \cite{kim1}
\begin{eqnarray} \label{vib1}
H_{\rm vib} &=& \sum_{i=1}^N \frac{p^2_i}{2M} + {M \nu^2 x_i^2 \over 2} + V(x_i) \, ,
\end{eqnarray}
where $M$ is the mass of a single atom. Moreover, $p_i$ is the momentum of particle $i$ in the direction of the cavity axis, $x_i$ denotes its distance from the trap centre and $V(x_i)$ describes corrections to the nearest harmonic trapping potential with phonon frequency $\nu$. To simplify this Hamiltonian, we now introduce bosonic phonon annihilation and creation operators $b_i$ and $b_i^\dagger$, 
\begin{eqnarray} \label{xipi}
b_i &\equiv & \sqrt{{M \nu \over 2 \hbar}} \, x_i + {\rm i} \, \sqrt{{1 \over 2 M \hbar \nu}} \, p_i \, , \nonumber \\
b_i^\dagger &\equiv & \sqrt{{M \nu \over 2 \hbar}} \, x_i - {\rm i} \, \sqrt{{1 \over 2 M \hbar \nu}} \, p_i \, , 
\end{eqnarray}
of particle $i$ with $[b_i, b_i^\dagger ] = 1$. When substituting these operators into Eq.~(\ref{vib1}), neglecting a constant term with no physical consequences and assuming, as an example, a cubic correction $V(x_i)$, the above Hamiltonian $H_{\rm vib}$ simplifies to
\begin{eqnarray} \label{last2}
H_{\rm vib} &=& \sum_{i=1}^N \hbar \nu \, b_i^\dagger b_i + \hbar \mu \, (b_i + b_i^\dagger)^3 \, ,
\end{eqnarray}
where $\mu$ is a constant. 

In order to enable us to analytically predict the dynamics of the mean phonon number $m$ and the phonon expectation value $\zeta$, we need to apply certain approximations. Firstly, we assume in the following that the trap is only slightly anharmonic, which implies $\mu \ll \nu$. Moreover, we assume that the trap confines the atoms within an area that is much smaller than the wavelength of the optical resonator. Achieving sub wavelength confinement of atomic particles inside an optical cavity is feasible with current technology \cite{Rempe2,Rempe3,Rempe4}. 

\subsection{Cooling stage} \label{walter}

As illustrated in Fig.~\ref{timeline}, each displacement stage is followed by a short resonant laser pulse with a (real) Rabi frequency $\Omega$. The experimental setup, which we need to consider when analysing the cooling stage, is shown in Fig.~\ref{setup}. The laser pulse changes the population of the electronic states of the trapped atoms, while affecting their vibrational degrees of freedom and placing photon excitations into the cavity field. To take this into account, we denote the ground state and the excited state of particle $i$ at position $r_i$ by $|0 \rangle_i$ and $|1 \rangle_i$, respectively, while $\sigma^+_i \equiv |1 \rangle_{ii} \langle 0|$ and $\sigma_i^- \equiv |0 \rangle_{ii} \langle 1|$. As before in the introduction, we denote the cavity photon annihilation operator by $c$. Moreover, $k_{\rm cav}$ is the wavenumber of the photons inside the resonator and $\omega_0$, $\omega_{\rm cav}$ and $g$ denote the atomic frequency, the cavity frequency and the (real) atom-cavity coupling constant, respectively. In the interaction picture with respect to the Hamiltonian $H_0$ of the cavity field and the atomic states,
\begin{eqnarray} \label{H0}
H_0 &=& \hbar \omega_0 \, c^\dagger c + \sum_{i=1}^N  \hbar \omega_0 \, \sigma_i^+ \sigma_i^- \, , 
\end{eqnarray}
the total Hamiltonian of the system can now be written as
\begin{eqnarray} \label{HI2first}
H_{\rm I} &=&  \sum_{i=1}^N {\hbar \Omega \over 2} \, \sigma_i^- + \hbar g \, \sin \big( k_{\rm cav} r_i \big) \, c \sigma_i^+ + {\rm H.c.} \nonumber \\
&& + \hbar \delta \, c^\dagger c + H_{\rm vib}
\end{eqnarray}
in the usual dipole and rotating wave approximations and with the atom-cavity detuning $\delta$ defined as
\begin{eqnarray} \label{delta}
\delta &=& \omega_{\rm cav} - \omega_0 \, .
\end{eqnarray}
To maximise the cooling process, this detuning should be equal to the phonon frequency $\delta  = \nu$, as indicated in Eq.~(\ref{delta}) and as illustrated in Fig.~\ref{Fig3}. For simplicity we neglect in the following direct dipole-dipole interactions between atomic particles. Since the applied laser field is in resonance with the atoms, transitions which do not involve the cavity field are not expected to change mean phonon numbers. Later, in Section \ref{peter}, we approximate the electronic states of the atoms by a stationary state. Dipole-dipole interactions might slightly change this stationary state but would not change the predictions in Section \ref{peter2} qualitatively.

Moreover, to maximise the coupling of the atomic motion to the cavity field mode, we assume in the following that the centre of the asymmetric trap $R$ coincides with a node of the resonator field, as illustrated in Fig.~\ref{setup}. This condition implies $\sin ( k_{\rm cav} R) = 0$. As mentioned already above, we assume in addition that the trap localises the atoms within an area much smaller than the relevant wavelength of their electronic transition. Taking this and the fact that $r_i = R+ x_i$ into account and applying the so-called Lamb-Dicke approximation yields 
\begin{eqnarray} \label{also}
\sin \big( k_{\rm cav} r_i \big) &=& \eta \, (b_i + b_i^\dagger)
\end{eqnarray}
up to terms which scale as $\eta^3$, where $\eta$ is the so-called Lamb-Dicke parameter and much smaller than one. Hence
\begin{eqnarray} \label{HI2first2}
H_{\rm I} &=&  \sum_{i=1}^N {\hbar \Omega \over 2} \, \sigma_i^- + \hbar \eta g \, (b_i + b_i^\dagger) \, c \sigma_i^+ + {\rm H.c.} \nonumber \\
&& + \hbar \delta \, c^\dagger c + H_{\rm vib} \, .
\end{eqnarray}
Finally, spontaneous emission from excited atoms and the leakage of photons through the cavity mirrors, is in the following taken into account by the quantum optical master equation \cite{Cirac2,kim1}
\begin{eqnarray} \label{master}
\dot{\rho}_{\rm I} &=& - {{\rm i} \over \hbar} \left[H_{\rm I},\rho_{\rm I} \right] + {\kappa \over 2} \left(2 c \rho_{\rm I} c^\dagger - c^\dagger c \rho_{\rm I} - \rho_{\rm I} c^\dagger c \right) \nonumber \\
&& + \sum_{i=1}^N {\Gamma \over 2} \left( 2 \sigma_i^- \rho_{\rm I} \sigma_i^+ - \sigma_i^+ \sigma_i^- \rho_{\rm I} - \rho_{\rm I} \sigma_i^+ \sigma_i^- \right) \, , ~~~~
\end{eqnarray}
where $\Gamma$ and $\kappa$ are the usual atom and cavity decay rates and where $\rho_{\rm I}$ denotes the density matrix of the atom-phonon-photon system in the interaction picture with respect to $H_0$ in Eq.~(\ref{H0}).

\section{Analysis of the cooling process} \label{analysis}

We now have all the theoretical tools needed to analyse the dynamics of the trapped atomic gas during the proposed cyclic two-stage cooling process. However before doing so, let us have a closer look at the most relevant expectation values of the experimental setup in Fig.~\ref{setup}. Usually, in laser cooling, there is only one slowly-evolving expectation value, namely the mean phonon number $m$ per particle (c.f.~Eq.~(\ref{notation})) \cite{Norah}. However, as we shall see below, the cooling scheme that we propose here has two slowly-evolving phonon expectation values, which play a crucial role in the cooling process. One is the mean phonon number $m$. The second variable is the many-particle phonon expectation value $\zeta$ which we introduced in Eq.~(\ref{notation}). In this section, we analyse the dynamics of both variables using standard quantum optical approximations.
 
\subsection{The physical meaning of $m$ and $\zeta$}

But before doing so, let us have a closer look at the physical meaning of $m$ and $\zeta$ in Eq.~(\ref{notation}). In this paper, we refer to $m$ and $\zeta$ as phonon expectation values, since they are expectation values of the bosonic operators $b_i$ in Eq.~(\ref{xipi}). If the particles were confined by a harmonic trapping potential, $m$ would be a direct measure of their mean vibrational energy. However, in an asymmetric trap, the the expectation value for the vibrational energy of the trapped atoms is given by $\langle H_{\rm vib} \rangle$ and no longer coincides automatically with $N \hbar \nu \, m$. Comparing Eqs.~(\ref{notation}) and (\ref{last2}), one can show that
\begin{eqnarray} \label{last2ddd}
\langle H_{\rm vib} \rangle &=& N \hbar \nu \, m + \sum_{i=1}^N \hbar \mu \, \left \langle (b_i + b_i^\dagger)^3 \right \rangle \, .
\end{eqnarray}
This shows that $\langle H_{\rm vib} \rangle = N \hbar \nu \, m$ only applies to a very good approximation when 
\begin{eqnarray} \label{last2xxx}
\mu \ll \nu ~~ &{\rm and/or}& ~~ m \ll 1 \, . 
\end{eqnarray}
In the remainder of this paper, we restrict ourselves to this parameter regime and consider only slightly asymmetric trapping potentials with $\mu \ll \nu$. Moreover, at the end of the cooling process which we analyse in this paper predict final phonon numbers $m_{\rm final}$ which are much smaller than one. Hence we can safely assume here that the atoms are eventually cooled down to very low temperatures as long as $m$ becomes eventually very small. In general, cooling only occurs as long as $\langle H_{\rm vib} \rangle$ decreases in time.

Next let us have a closer look at the physical meaning of the phonon expectation value $\zeta$. Combining Eqs.~(\ref{notation}) and (\ref{xipi}), one can show that
\begin{eqnarray} \label{notation2}
\zeta &=& {1 \over N(N-1)} \sum_{i=1}^N  \sum_{j \neq i}  {M \nu \over 2 \hbar} \, \langle x_i x_j \rangle +  {1 \over 2 \hbar M \nu} \, \langle p_i p_j \rangle ~~~~~~
\end{eqnarray}
without any approximations. Assuming that all particles move independently and that the atomic confinement is so strong that there is no flux of particles out of the trap and applying semiclassical standard approximations, this implies that 
\begin{eqnarray} \label{simple}
\langle x_i x_j \rangle &=& \langle x_i \rangle \langle x_j \rangle \, , \nonumber \\
\langle p_i p_j \rangle &=& \langle p_i \rangle \langle p_j \rangle = 0 \, . 
\end{eqnarray}
Combining these two observations with Eq.~(\ref{notation2}) shows
\begin{eqnarray} \label{zetazz}
\zeta &=& {M \nu \over 2 \hbar} \, \langle x \rangle^2
\end{eqnarray}
to a very good approximation, where $\langle x \rangle$ denotes the mean distance of the trapped particles from the centre of the trap.  When this equation applies, $\zeta$ is always positive and a measure for the distance of the atomic gas from the centre of the trap. For example, $\langle x \rangle =0$ implies $\zeta = 0$ and the accumulation of the atomic particles around the centre of the trap.

Having a closer look at Eq.~(\ref{notation2}), which is exact, we see that the classical analog of the phonon expectation value $\zeta$ is in general different from zero. For example, if the symmetry of the trap is broken and many more particles accumulate on one side of the centre of the trap than on the other, then the first term in Eq.~(\ref{notation2}) contains significantly more positive than negative contributions. If the particles move moreover in a synchronised fashion, such that the momenta of a majority of particles have the same sign, then the second term in Eq.~(\ref{notation2}) increases $\zeta$ even further. In the following, we ignore the possibility of synchronised motion and estimate $\zeta$ at the end of every displacement stage using Eq.~(\ref{zetazz}), thereby potentially underestimating the efficiency of the proposed cooling scheme. However, as we shall see below, at the end of every cooling stage, we have $\zeta = 0$ which suggests the absence of synchronisation due to the common coupling of the particles to the cavity field. As one would expect for independently and freely moving particles, we then obtain a positive value for $\zeta$ at the end of each displacement stage which increases as the trap asymmetry increases.

Notice also that $\zeta$ can only be defined when there are at least two particles in the trap. For only one particle, Eq.~(\ref{zetazz}) does not apply and we always have $\zeta \equiv 0$ (c.f.~Eq.~(\ref{notation})). This is why the cooling process, which we describe in this paper, is qualitatively different from single-particle laser-cooling schemes. Single-particle cavity cooling schemes cannot take advantage of a non-zero $\zeta$ as an additional resource.

\subsection{Displacement stage} \label{displ}

As we shall see below, at the end of every cooling stage, and therefore also at the beginning of every displacement stage, we have $\zeta = 0$, while the average mean phonon number $m$ is in general non-zero. The main purpose of the displacement stage is to increase $\zeta$, while keeping $m$ essentially the same. In this subsection, we show that this task can be achieved by allowing the particles oscillate freely and independently within their asymmetric trapping potential. To do so, we use the conservation of the vibrational energy  and Eq.~(\ref{zetazz}) to estimate the dynamics of $\zeta$ and $m$. In the absence of any laser driving and other quantum effects, the atomic particles move according to the laws of classical mechanics and there is nothing quantum about their motion.

Unfortunately, analysing the dynamics of particles inside an asymmetric trapping potential, even classically, is not straightforward. In this subsection, we therefore employ an approximation. To determine the average position $x(t)$ of a single trapped particle after a sufficiently long time $t$, we assume in the following that 
\begin{eqnarray} \label{fin2}
x(t) &=& {x_{\rm min} + x_{\rm max} \over 2}
\end{eqnarray}
to a very good approximation. Here $x_{\rm min}$ and $x_{\rm max}$ are the return points of the free oscillations of the atom. These two positions can be calculated relatively easily, since they correspond to points of vanishing momentum $p$. Combining Eqs.~(\ref{xipi}) and (\ref{last2}) and setting $p=0$, we find that 
\begin{eqnarray} \label{funny}
H_{\rm vib} &=& {M \nu^2 \over 2} \, x^2 + \hbar \mu \left( {2 M \nu \over \hbar} \right)^{3/2} x^3 ~~~
\end{eqnarray}
for $x = x_{\rm min}$ and $x = x_{\rm max}$. Moreover, we know that the vibrational energy of the trapped particle $H_{\rm vib}$, which is conserved, equals 
\begin{eqnarray} \label{funny2}
H_{\rm vib}(0) &=& \hbar \nu \, m (0)
\end{eqnarray}
at the beginning of the displacement stage, since $\zeta$ and therefore also $x$ are both zero at $t=0$. Combining Eqs.~(\ref{fin2})-(\ref{funny2}) yields  
\begin{eqnarray} \label{19}
x(t) &=& - {4 \mu \over \nu} \left( {2 \hbar \over M \nu} \right)^{1/2} \, m(0)
\end{eqnarray}
up to first order in $\mu/\nu$. Substituting this result into Eq.~(\ref{zetazz}), we finally see that
\begin{eqnarray} \label{notation56}
\zeta (t) &=& \left( {4 \mu \over \nu} \right)^2 \, m(0)^2 
\end{eqnarray}
to a good approximation. As one would expect, $\zeta(t)$ is positive and non-zero at the end of every displacement stage, unless $H_{\rm vib}$ tends to zero. Moreover, $\zeta$ in Eq.~(\ref{notation56}) increases rapidly as the vibrational energy, i.e.~$m(0)$, of the atomic particles increases. The hotter the atoms are, the further away they accumulate on average from the centre of the trap.

Combining Eqs.~(\ref{xipi}), (\ref{last2}), (\ref{zetazz}) and (\ref{notation56}), we find that the vibrational energy of a single atomic particle equals 
\begin{eqnarray}
H_{\rm vib} (t) &=& \hbar \nu \, m(t) + 8 \hbar \mu \, \zeta(t)^{3/2} 
\end{eqnarray}
at the end of the displacement stage. Conservation of the vibrational energy during the free evolution of the trapped particle implies that this expression is the same as $H_{\rm vib}(0)$ in Eq.~(\ref{funny2}). Using this and Eq.~(\ref{notation56}), one can now show that 
\begin{eqnarray}
m (t) &=& m(0) \left[ 1 - 2 \left( {4\mu \over \nu} \right)^4 m(0)^2 \right] 
\end{eqnarray}
to a very good approximation. For sufficiently small trap asymmetries $\mu/\nu$, the average mean phonon number $m$ remains essentially the same.

\subsection{Cooling stage} \label{sevi}

Next we analyse the dynamics of the expectation values $m$ and $\zeta$ when the cooling laser is turned on. To do so, we first simplify the full master equation in Eq.~(\ref{master}) to an effective master equation that has many similarities with the master equation of the idealised toy model in Section \ref{introduction}. Afterwards we analyse its effective time evolution. It is shown that the mean number of phonons per particle $m$ drops very rapidly by an amount which equals the initial value of the phonon expectation value $\zeta$. In other words, the dynamics of the experimental setup are analogous to the dynamics described in Section \ref{introduction}, even when off-resonant interaction terms in the system Hamiltonian, additional decay channels and the asymmetry of the trap are taken into account. What we find is that the change of the mean phonon number $m$ depends only on the sign of $\zeta$. As long as $\zeta $ is positive, we predict a reduction of $m$ with a collectively enhanced cooling rate, as described in Eq.~(\ref{msmall}).

\subsubsection{An effective master equation} \label{peter}

As illustrated in Fig.~\ref{Fig3}, the role of the atoms is to bridge the energy gap between phonons and cavity photons, thereby helping to create an effective resonant direct exchange interaction between these different types of particles. Given the parameters of typical atom-cavity experiments, where \cite{footnote}
\begin{eqnarray} \label{condi}
\Omega , \, \Gamma &\gg & \nu, \, \kappa, \, \delta, \, g \, , 
\end{eqnarray}
the conversion is usually relatively inefficient, even when it is collectively enhanced, while the electronic states of the atoms evolve on a relatively fast time scale. In the following, we take this into account and approximate the state of the atoms by the stationary state, which they would assume in the absence of any off-resonant coupling to the cavity field and phonon modes. More concretely, we assume in the following that the atomic density matrix $\rho_{\rm A}(t)$ evolves in the interaction picture according to the master equation
\begin{eqnarray} \label{master2}
\dot{\rho}_{\rm A} &=& - {{\rm i} \Omega \over 2} \left[ \sigma_i^- + \sigma_i^- , \rho_{\rm A} \right] \nonumber \\
&& + \sum_{i=1}^N {\Gamma \over 2} \left( 2 \sigma_i^- \rho_{\rm A} \sigma_i^+ - \sigma_i^+ \sigma_i^- \rho_{\rm A} - \rho_{\rm A} \sigma_i^+ \sigma_i^- \right) ~~~~~
\end{eqnarray}
and approximate the state of the atoms by the stationary state $\rho_{\rm A}^{\rm ss}$ with $\dot{\rho}_{\rm A} = 0$. 

In other words, we assume in the following that the density matrix $\rho_{\rm I} (t)$ can always be written as
\begin{eqnarray} \label{master2zzz}
\rho_{\rm I} (t) &=& \rho_{\rm BC} (t) \otimes \rho_{\rm A}^{\rm ss} \, , 
\end{eqnarray}
where $\rho_{\rm BC}(t)$ denotes the density matrix of phonons and cavity photons at time $t$. If the atoms reach their stationary state $\rho_{\rm A}^{\rm ss}$ very rapidly without changing the statistical properties of the phonons and photons in the system, the master equation (\ref{master}) can be used to show that, for a small time interval $\Delta t$,
\begin{eqnarray} \label{master3}
\dot{\rho}_{\rm BC} (t + \Delta t) &=& {\rm Tr}_{\rm A} \left( \rho_{\rm I} (t) + \dot{\rho}_{\rm I} (t) \Delta t \right) \, ,  
\end{eqnarray}
where ${\rm Tr}_{\rm A}( \cdot )$ denotes the trace operation over the state space of the electronic states of the atoms. Combining this equation with Eqs.~(\ref{master}) and (\ref{master2}), we find that the time derivative $\rho_{\rm BC} (t)$ equals 
\begin{eqnarray} \label{master4}
\dot{\rho}_{\rm BC} (t) &=& {\rm Tr}_{\rm A} \left(\dot{\rho}_{\rm I} (t) \right) \, .
\end{eqnarray}
Hence the density matrix $\rho_{\rm BC}$ evolves effectively according to the master equation
\begin{eqnarray} \label{master3}
\dot{\rho}_{\rm BC} &=&  {\rm Tr}_{\rm A} \left( - {{\rm i} \over \hbar} \left[H_{\rm I} , \rho_{\rm BC} \otimes  \rho_{\rm A}^{\rm ss} \right] \right) \nonumber \\
&& + {\kappa \over 2} \left(2 c \, \rho_{\rm BC} \, c^\dagger - c^\dagger c \, \rho_{\rm BC} - \rho_{\rm BC} \, c^\dagger c \right) \, . ~~~~
\end{eqnarray}
This equation is of exactly the same form as Eq.~(\ref{1}) but, in contrast to Eq.~(\ref{Heff}), the effective phonon-photon interaction Hamiltonian $H_{\rm eff}$ is now given by 
\begin{eqnarray} \label{HI2first3}
H_{\rm eff} &=&  \sum_{i=1}^N \hbar \eta g \, _i \langle 0 | \rho_{\rm A}^{\rm ss} |1 \rangle_i \left( b_i + b_i^\dagger \right) c + {\rm H.c.} \nonumber \\
&& + \hbar \delta \, c^\dagger c + H_{\rm vib} \, ,
\end{eqnarray}
where $|0 \rangle_i$ denotes the ground state of atom $i$, while $|1 \rangle_i$ denotes its excited state. Proceeding as described in App.~\ref{appA1}, analysing the atomic dynamics and calculating the corresponding stationary state in the presence of resonant laser driving and spontaneous photon emission, we finally find that
\begin{eqnarray} \label{HI2first4}
H_{\rm eff} &=& \sum_{i=1}^N {\rm i} \hbar \eta g_{\rm eff} \left( b_i + b_i^\dagger \right) \left( c - c^\dagger \right) +  \hbar \nu \, b_i^\dagger b_i \nonumber \\
&& + \hbar \mu \, \left( b_i + b_i^\dagger \right)^3 + \hbar \delta \, c^\dagger c  
\end{eqnarray}
with the effective coupling constant $g_{\rm eff} $ given by
\begin{eqnarray} \label{mus3}
g_{\rm eff} &=& - {g \Gamma \Omega \over \Gamma^2 + 2 \Omega^2} \, .
\end{eqnarray}
This Hamiltonian has many similarities with the toy model Hamiltonian $H_{\rm eff}$ in Eq.~(\ref{Heff}). Moreover, the master equation of the phonon and cavity photon system for the experimental setup in Fig.~\ref{setup} is the same as the toy model master equation in Eq.~(\ref{1}). As we shall see below, for positive atom-cavity detunings $\delta $, the phonon dynamics of the proposed cooling scheme is essentially the same as the phonon dynamics described in Section \ref{introduction}. To show that this is indeed the case, we now analyse the dynamics of $m$ and $\zeta$ with all additional coupling terms, detunings and decay channels taken into account. 

\subsubsection{Effective cooling equations} \label{peter2}

First we have a closer look at the dynamics of the slowly evolving variables $m$ an $\zeta$. Using the above derived effective system dynamics (c.f.~Eqs.~(\ref{master3}) and (\ref{HI2first4})) and proceeding as described in App.~\ref{xcoherences}, one can now show that
\begin{eqnarray} \label{mbetadot}
\dot{m} &=& \eta g_{\rm eff} \, x_{33} + 3 \mu \, y_{232} \, , \nonumber \\
\dot \zeta &=& \eta g_{\rm eff} \, x_{33} + 3 \mu \, \tilde y_{223} 
\end{eqnarray}
without any approximations. The definition of the $x$, $y$ and $\tilde y$ variables on the right hand side of this equation can be found in Eq.~(\ref{notation500}) in App.~\ref{lewis}. Here $x_{33}$ and $y_{232}$ are single-particle expectation values, while $\tilde y_{223}$ denotes a many-particle phonon expectation value. As shown in App.~\ref{xcoherences}, the variables $x_{33}$, $y_{232}$ an $\tilde y_{223}$ evolve on a much faster time scale than $m$ and $\zeta$ when
\begin{eqnarray} \label{condi2}
N \eta g_{\rm eff}, \, \nu, \, \kappa, \, \delta  &\gg & \eta g_{\rm eff} , \,  \mu \, . 
\end{eqnarray}
Taking this into account, proceeding as described in App.~\ref{xcoherences2} and eliminating these variables adiabatically from the system dynamics, yields a closed set of two effective cooling equations. In analogy to Eq.~(\ref{rates}), we find that
\begin{eqnarray} \label{erna}
\dot m = \dot \zeta &=& - A_N \, \zeta 
\end{eqnarray}
with the collective rate $A_N$ given by
\begin{eqnarray} \label{fun}
A_N = {N (8\eta g \Gamma \Omega)^2 \kappa \delta \nu \over \left[ \kappa^4 + 16 (\delta^2 - \nu^2)^2 + 8 \kappa^2 (\delta^2 + \nu^2) \right] (\Gamma^2 + 2 \Omega^2)^2} \, .  \nonumber \\
\end{eqnarray}
The factor $N$ in Eq.~(\ref{fun}) shows that the time evolution of the mean phonon number $m$ is collectively enhanced. The system succeeds in taking advantage of the simultaneous coupling the atoms to the cavity field.

The effective rate equations in Eq.~(\ref{erna}) are formally the same as the rate equations in Eq.~(\ref{rates}) and can be solved easily analogously. Suppose the atom-cavity detuning $\delta$ and the initial phonon expectation value $\zeta (0)$ are both positive and the cooling laser is turned on at $t=0$ for a time $t \gg 1/A_N$. Then $A_N$ is positive and $m(t)$ and $\zeta(t)$ are again to a very good approximation given by Eq.~(\ref{msmall}). The cooling laser reduces the mean phonon number $m$ by $\zeta(0)$. Playing the role of a resource, the variable $\zeta$ itself vanishes in the process which implies $\langle x(t) \rangle = 0$ at the end of the cooling stage (c.f.~Eq.~(\ref{zetazz})). Under the action of the laser cooling, the atomic gas returns on average to the centre of the trap. 

One can easily see from Eq.~(\ref{fun}) that the cooling laser reduces the mean phonon number $m$ most rapidly when $\delta = \nu$. In this case, the effective cooling rate $A_N$ simplifies to
\begin{eqnarray} \label{fun2}
A_N &=& {N (8 \eta g \Gamma \Omega \nu)^2 \over \kappa (\kappa^2 + 16 \nu^2) (\Gamma^2 + 2 \Omega^2)^2} \, . 
\end{eqnarray}
This is not unexpected. When $\delta  = \nu$, the most likely transition in the dynamics of the system is the de-excitation of a particle accompanied by the annihilation of a phonon and the creation of a cavity photon \cite{kim1}. Once the photon leaks out of the resonator, the phonon is permanently lost. Moreover, heating transitions, which involve the simultaneous excitation of the cavity and the creation of a phonon, are most unlikely in this case. However these heating and cooling transitions only affect the mean number of phonons $m$ on a time scale much longer than $1/A_N$. For the cavity-mediated collective cooling scheme, which we propose here, they remain negligible. Instead, the dynamics of the cooling process is dominated by the collective effects which we illustrated in the Introduction with the help of a relatively simple toy model.

\subsubsection{Some physical intuition}

In Section \ref{walter}, we assume that the atoms are strongly confined and that the centre of their trapping potential coincides with a node of the cavity field (c.f.~Fig.~\ref{setup}). In this case, the above described cooling process can be understood relatively easily also intuitively. As illustrated in Fig.~\ref{Fig3}, the applied laser field drives the atoms resonantly. Changing $m$ therefore requires off-resonant interactions with the cavity field. For $\zeta = 0$, the trapped particles experience equal amounts of cooling and heating, since their atom-cavity interaction becomes zero, and $m$ remains constant in this case. However, for $\zeta \neq 0$ the dynamics of the system during cooling stages becomes less trivial. Cooling transitions move the atoms closer to the centre of the trap and reduce $\zeta$. Analogously, heating transitions move the atoms further away and increases $\zeta$. The dynamics of hotter particles is hence more intense. Eventually, all particles reach the centre of the trap which corresponds to a stable fixed point of their dynamics and where $\zeta =0$. The overall effect of the transfer of the atoms, from their initial position to the centre of the trap, is a loss of vibrational energy (c.f.~Eq.~(\ref{msmall})). To re-initiate the cooling process, the atoms need to be moved again away from the cavity node. Another displacement stage is needed.

\section{The final mean phonon number} \label{estimate}

To achieve very low temperatures, we propose to alternate cooling stages with displacement stages as illustrated in Fig.~\ref{timeline}. During each displacement stage, the non-interacting atomic gas evolves freely and naturally accumulates a non-zero phonon coherence $\zeta$ without increasing its mean phonon number $m$. During each cooling stage, $\zeta$ tends to zero, while $m$ decreases by the same amount in the process (c.f.~Eq.~(\ref{msmall})). The hotter the atoms are, the larger their initial phonon coherence $\zeta$ (c.f.~Eq.~(\ref{notation56})). Eventually, both expectations values, $m$ and $\zeta$, become too small to dominate the dynamics of the system during cooling stages (c.f.~Eq.~(\ref{erna})). At this point, other cooling and heating processes, which occur only on a time scale given by $\eta^2$, are no longer negligible. These transitions have previously been analysed in detail in Ref.~\cite{kim1} for example. 

To get at least an estimate for the final mean phonon number $m_{\rm final} $ of the proposed cooling process, we now combine Eqs.~(\ref{notation56}) and (\ref{erna}) with Eq.~(39) in Ref.~\cite{kim1}. Doing so we find that the time derivative of $m$ at the beginning of the next cooling stage equals 
\begin{eqnarray}
\dot m &=& - {16 \mu^2 A_N \over \nu^2} \, m^2 - \gamma_c \, m + c 
\end{eqnarray}
also to a very good approximation, where $\gamma_c$ and $c$ are two constants of order $\eta^2$. Assuming that the cooling process ends, when $\dot m = 0$, we find that   
\begin{eqnarray} \label{mubash}
m_{\rm final} &=& {\nu  \over 4 \mu} \left( {c \over  A_N} \right)^{1/2}  
\end{eqnarray}
to a very good approximation. Most importantly, the right hand side of this equation scales as $1/N^{1/2}$ and tends to zero in the many-particle limit. For example, if we choose $\mu/\nu = 0.01$ and $N=10^6$, we have $\nu/(4 \mu N^{1/2}) = 0.025$ which is much smaller than one. The cooling process, which we propose in this paper, therefore has the potential to cool an atomic gas to much lower temperatures than other laser cooling schemes. 

For more traditional cavity-mediated laser cooling schemes, the final temperature of the atoms scales essentially as $(\kappa/\nu)^2$, which is independent of the number $N$ of atoms inside the trap. For example, substituting $c$ from Eq.~(41) in Ref.~\cite{kim1} into Eq.~(\ref{mubash}) yields 
\begin{eqnarray} \label{mubash2}
m_{\rm final}  &=& {\kappa \over 16 \mu N^{1/2}} 
\end{eqnarray}
when $\delta = \nu$ and $\Omega, \nu \ll \kappa, \Gamma$. As one would expect, collective cooling of an atomic gas to very low temperatures with the cavity-mediated laser cooling scheme which we propose here requires a not-too-small asymmetry of the trapping potential. Moreover the particles need to be strongly confined and the cavity decay rate $\kappa$ should be relatively small. Although the proposed cooling scheme might not require a cavity, which operates in the strong-coupling regime, the cavity should not be too bad either, as one would expect.

\section{Conclusions}  \label{conclusions}

This paper proposes a cyclic two-stage laser cooling scheme to transfer a non-interacting atomic gas to very low temperatures. The corresponding experimental setup, in which the atoms are confined to an asymmetric trapping potential inside an optical resonator, is shown in Fig.~\ref{setup}. The role of the cavity and the external laser field is to remove vibrational energy from the atoms and to release it in the form of photons into the environment. The role of the trap asymmetry is to create a positive phonon coherence $\zeta$ between laser interactions without affecting the free energy of the trapped particles. Our calculations assume that the atoms are strongly confined within a wavelength of the resonator. Moreover we assume that the asymmetry of the trap can be described by a cubic term with a constant $\mu$, which is much smaller than the phonon frequency $\nu$. Although the parameter regime, which we consider in this paper, is relatively standard for cavity-mediated laser-cooling (c.f.~Eqs.~(\ref{last2xxx}), (\ref{condi}) and (\ref{condi2})), we find that the usual temperature limits do not apply. Instead, the final vibrational energy of each atom scales as one over the square root of the total number of particles $N$ (cf.~Eq.~(\ref{mubash2})) which vanishes in the infinitely-many particle limit.

One drawback of the proposed cooling scheme is that its cools the particles in only one dimension. However, as long as there is an energy exchange between different directions of motion, draining kinetic energy out of the $x$ direction eventually also reduces the motion of the atoms in the $y$ and in the $z$ direction. Another drawback of the proposed cooling scheme is that experimental parameters need to be chosen relatively carefully. Eq.~(\ref{mubash2}) shows that the particle number $N$ should be relatively large, while the cavity decay rate $\kappa$ and the trap asymmetry $\mu$ should not be too large and too small, respectively, in order to outperform standard laser cooling schemes. Nevertheless, the non-destructive cooling mechanism, which we propose in this paper, could become an important tool for quantum technological applications. \\[0.5cm]
{\em Acknowledgments.} AB would like to dedicate this paper to her dear and very highly-regarded colleague, Danny Segal, who she met and interacted with during her time at Imperial College London. Moreover, OK and AB thank the UK Engineering and Physical Sciences Research Council EPSRC (Grant Ref.~EP/H048901/1) and PD thanks the British Council and DST Government of India for financial support. We also acknowledge inspiring discussions with P.~Grangier and thank him for his constructive feedback on the manuscript.

\appendix
\section{Derivation of effective cooling equations for a ideal model} \label{App}

Using Eq.~(\ref{1}), one can show that the time derivative of the expectation value $\langle A \rangle$ of an observable $A$ of our idealised cooling scheme equals
\begin{eqnarray} \label{tracey}
\langle \dot{A} \rangle &=& - {{\rm i} \over \hbar} \langle \left[ A, H_{\rm ideal} \right] \rangle + {\kappa \over 2} \left \langle \big[c^\dagger , A \big] c + c^\dagger \big[A , c \big] \right \rangle . ~~~~~~
\end{eqnarray}
For the parameter regime given in Eq.~(\ref{5}), the dynamics of $m$ and $\zeta$ is therefore to a very good approximation given by the closed set of rate equations
\begin{eqnarray} \label{rates2}
\dot{m} = \dot{\zeta} &=& J \, x \, , \nonumber \\
\dot{x} &=& - 2 N J \, \zeta - {\kappa \over 2} \, x 
\end{eqnarray}
with the additional multi-particle quantum coherence $x$ defined as 
\begin{eqnarray} \label{notation3}
x &\equiv & {{\rm i} \over N} \sum_{i=1}^N  \langle b_i c^\dagger - b_i^\dagger c \rangle \, .
\end{eqnarray}
The only approximation applied here is to neglect terms which scale as $J$ in the time derivative of $x$. Since $N \gg 1$, the variable $x$ evolves on a much faster time scale than the expectation values $m$ and $\zeta$. Hence $x$ can be adiabatically eliminated from the system dynamics. Doing so and assuming $\dot{x} = 0$, we find that 
\begin{eqnarray} \label{xxx}
x &=& - {4 N J \over \kappa} \, \zeta
\end{eqnarray}
to a very good approximation. Substituting this result into the differential equations for $m$ and $\zeta$ in Eq.~(\ref{rates2}) eventually yields the closed set of two effective cooling equations in Eq.~(\ref{rates}).

\section{Derivation of effective cooling equations for the proposed cooling scheme} \label{appA}

As pointed out already in Section \ref{sevi}, the purpose of this appendix is to derive a closed set of rate equations, for the time evolution of the expectation values $m$ and $\zeta$ which we introduced in Eq.~(\ref{notation2}). To do so, we first analyse the zeroth order dynamics of the atoms in the presence of resonant laser driving and spontaneous photon emission. Afterwards, we introduce further expectation values which we need to consider in addition to the slowly evolving expectation values $\zeta$ an $m$. Afterwards, we have a closer look at the time derivatives of these relatively rapidly evolving variables and calculate them with the help of an adiabatic elimination technique which is routinely used in quantum optics. The following results apply to a very goo approximation, as long as the system parameters obey the inequalities summarised in Eq.~(\ref{condi}).

\subsection{Atomic dynamics} \label{appA1}

First, we calculate the expectation value $_i \langle 0 | \rho_{\rm A}^{\rm ss} |1 \rangle_i$ in Eq.~(\ref{HI2first3}) by analysing the dynamics of the electronic states of the atoms for the parameter regime given in Eq.~(\ref{condi}). Since $\rho_{\rm A}$ evolves according to the effective atomic master equation Eq.~(\ref{master2}), the time derivative of the expectation value $\langle A \rangle$ of any atomic observable $A$ equals
\begin{eqnarray} \label{traceyx}
\langle \dot{A} \rangle &=& - {{\rm i} \Omega \over 2}  \left \langle \left[ A, \sigma_i^- + \sigma_i^- \right] \right \rangle \nonumber \\
&& + \sum_{i=1}^N {\Gamma \over 2} \, \left \langle \left[ \sigma_i^+ , A \right] \sigma_i^- + \sigma_i^+ \left[A , \sigma_i^- \right]  \right \rangle \, . ~~~~
\end{eqnarray}
To obtain a closed set of rate equations, we consider the operators
\begin{eqnarray} \label{Sevis}
\left( \Sigma_1^{(i)}, \Sigma_2^{(i)}, \Sigma_3^{(i)} \right) = \left( \sigma_i^+ \sigma_i^-, \sigma_i^- + \sigma_i^+ , {\rm i} \left( \sigma_i^- - \sigma_i^+ \right) \right) ~~~~
\end{eqnarray}
with the commutator relations
\begin{eqnarray} \label{commsx}
&& \hspace*{-0.5cm} \left[ \Sigma_1^{(i)}, \Sigma_2^{(i)} \right] = {\rm i} \Sigma_3^{(i)} \, , ~~ 
\left[ \Sigma_1^{(i)}, \Sigma_3^{(i)} \right] = - {\rm i} \Sigma_2^{(i)} \, ,  \nonumber \\ 
&& \hspace*{-0.5cm} \left[ \Sigma_2^{(i)}, \Sigma_3^{(i)} \right] = - 2 {\rm i} \left( 1 - 2 \Sigma_1^{(i)} \right) 
\end{eqnarray}
and denote their respective expectation values by $s_1$, $s_2$ and $s_3$. Combining Eqs.~(\ref{traceyx})-(\ref{commsx}), one can then show that
\begin{eqnarray}
&& \hspace*{-0.5cm} \dot s_1^{(i)} = {\Omega \over 2} \, s_3^{(i)} - \Gamma \, s_1^{(i)} \, , ~~ 
\dot s_2^{(i)} = - {\Gamma \over 2} \, s_2^{(i)} \, , ~~ \nonumber \\
&& \hspace*{-0.5cm} \dot s_3^{(i)} = \Omega \left(1 - 2 s_1^{(i)} \right) - {\Gamma \over 2} \, s_3^{(i)} \, .
\end{eqnarray}
Consequently, the expectation values of the stationary atomic state $\rho_{\rm A}^{\rm ss}$ with $\dot{s}_1^{(i)} = \dot{s}_2^{(i)} = \dot{s}_3^{(i)} = 0$, are given by
\begin{eqnarray} \label{mus}
\Big\{ s_1^{\rm ss},s_2^{\rm ss},s_3^{\rm ss} \Big\} &=& \left\{ {\Omega^2 \over \Gamma^2 + 2 \Omega^2} , 0, {2 \Gamma \Omega \over \Gamma^2 + 2 \Omega^2} \right\} 
\end{eqnarray}
for all particles $i$. This implies
\begin{eqnarray} \label{mus2}
_i \langle 0 | \rho_{\rm A}^{\rm ss} |1 \rangle_i &=& - {{\rm i} \Gamma \Omega \over \Gamma^2 + 2 \Omega^2} 
\end{eqnarray}
for all particles $i$, since $_i \langle 0 | \rho_{\rm A}^{\rm ss} |1 \rangle_i = ( s_2^{\rm ss} - {\rm i} \, s_3^{\rm ss} )/2$. Substituting the result into Eq.~(\ref{HI2first3}) yields the final the effective Hamiltonian $H_{\rm eff}$ in Eq.~(\ref{HI2first4}).

\subsection{Relevant expectation values} \label{lewis}

In analogy to Eq.~(\ref{Sevis}), we now introduce the operators $B_j^{(i)}$ and $C_k$ such that 
\begin{eqnarray} \label{oops}
\left( B_1^{(i)}, B_2^{(i)}, B_3^{(i)} \right) &=& \left( b_i^\dagger b_i, b_i + b_i^\dagger , {\rm i} \left( b_i - b_i^\dagger \right) \right) \, , \nonumber \\
\left( C_1, C_2, C_3 \right) &=& \left( c^\dagger c, c + c^\dagger , {\rm i} \left( c - c^\dagger \right) \right) \, .
\end{eqnarray}
Using this notation, the quantum coherence $\zeta$ and the average mean phonon number $m$ can be written as
\begin{eqnarray} \label{notationfinal}
\zeta &=& {1 \over 4N(N-1)} \sum_{i=1}^N \sum_{j \neq i} \left \langle B_2^{(i)} B_2^{(j)} + B_3^{(i)} B_3^{(j)} \right \rangle \, ,  \nonumber \\
m &=& {1 \over N} \sum_{i=1}^N \left \langle B_1^{(i)} \right \rangle \, . 
\end{eqnarray}
As we shall see below, in order to analyse the time evolution of these two variables we moreover need to consider the dynamics of the averaged expectation values 
\begin{eqnarray} \label{notation500}
x_{ab} &\equiv & {1 \over N} \sum_{i = 1}^N  \left \langle B_a^{(i)} C_b \right \rangle  \, , \nonumber \\
y_{abc} &\equiv & {1 \over N} \sum_{i = 1}^N  \left \langle B_a^{(i)}  B_b^{(i)}  B_c^{(i)}  \right \rangle  \, , \nonumber \\
\tilde y_{abc} &\equiv & {1 \over N(N-1)} \sum_{i = 1}^N \sum_{j \neq i} \left \langle B_a^{(i)}  B_b^{(i)}  B_c^{(j)}  \right \rangle \, , \nonumber \\
\tilde z_{ab} &\equiv & {1 \over N(N-1)} \sum_{i = 1}^N \sum_{j \neq i} \left \langle B_a^{(i)}  B_b^{(j)} \right \rangle \, . ~~~
\end{eqnarray}
All of the above expectation values are macroscopic quantities. They are independent of the number $N$ of atoms in the trap in the large $N$ limit.

\subsection{Time evolution} \label{xcoherences}

Next we proceed as described in Section \ref{introduction} and use the effective Hamiltonian $H_{\rm eff}$ in Eq.~(\ref{HI2first4}) together with the master equation in Eq.~(\ref{1}) to derive effective rate equations for the expectation values $m$ and $\zeta$ and the expectation values which we introduced in the previous subsection.  During these calculations, we make extensive use of the commutator relations 
\begin{eqnarray} \label{comms}
&& \hspace*{-0.5cm} \left[ B^{(i)}_1,B^{(i)}_2 \right] = {\rm i} B^{(i)}_3 \, , ~~
\left[ B^{(i)}_1,B^{(i)}_3 \right] = - {\rm i} B^{(i)}_2 \, , \nonumber \\
&& \hspace*{-0.5cm}  \left[ B^{(i)}_2,B^{(i)}_3 \right] = -2 {\rm i} \, , ~~ 
[ C_1,C_2] = {\rm i} C_3 \, , \nonumber \\
&& \hspace*{-0.5cm} [ C_1,C_3] = - {\rm i} C_2 \, , ~~
[ C_2,C_3] = -2 {\rm i} 
\end{eqnarray}
which we summarise here for convenience. Moreover, using Eq.~(\ref{oops}), Eq.~(\ref{HI2first2}) simplifies to
\begin{eqnarray} \label{HI4}
H_{\rm eff} &=& \sum_{i=1}^N \hbar \eta g_{\rm eff} \, B_2^{(i)} C_3 +  \hbar \nu \, B_1^{(i)} + \hbar \mu \, B_2^{(i) 3} \nonumber \\
&& + \hbar \delta \, C_1 \, . 
\end{eqnarray}
In the following, we combine Eq.~(\ref{tracey}) (but with $H_{\rm ideal}$ replaced by $H_{\rm eff}$) with Eqs.~(\ref{comms}) and (\ref{HI4}) to obtain the time derivatives in Eq.~(\ref{mbetadot}) which describe the relatively slow dynamics of $m$ and $\zeta$.

Given the parameter regime in Eq.~(\ref{condi2}), the expectation values $x_{ab}$ which we defined in Eq.~(\ref{notation500}) evolve on a much faster time scale than $m$ and $\zeta$. Taking this into account and neglecting terms which scale as $\eta g_{\rm eff}$ and $\mu$, we find that the single-particle expectation values $x_{ab}$ evolve to a very good approximation according to the differential equations
\begin{eqnarray} \label{ij0}
\dot x_{22} &=& - 2 N \eta g_{\rm eff} \, \tilde z_{22} - \nu \, x_{32} - \delta \, x_{23} - {\kappa \over 2} \, x_{22} \, , \nonumber \\
\dot x_{23} &=& - \nu \, x_{33} + \delta \, x_{22} - {\kappa \over 2} \, x_{23} \, , \nonumber \\
\dot x_{32} &=& - 2 N \eta g_{\rm eff} \, \tilde z_{23} + \nu \, x_{22} - \delta \, x_{33} - {\kappa \over 2} \, x_{32} \, , \nonumber \\
\dot x_{33} &=& \nu \, x_{23} + \delta \, x_{32} - {\kappa \over 2} \, x_{33} \, . 
\end{eqnarray}
Moreover, the time derivative of the two-particle quantum coherences $\tilde z_{ab}$ are given by
\begin{eqnarray} \label{ij1}
\dot {\tilde z}_{22} &=& - 2 \nu \, \tilde z_{23} \, , \nonumber \\
\dot {\tilde z}_{23} = \dot {\tilde z}_{32} &=& \nu \left( {\tilde z}_{22} - {\tilde z}_{33} \right) \, , \nonumber \\
\dot {\tilde z}_{33} &=& 2 \nu \, {\tilde z}_{23} \, ,
\end{eqnarray}
if we neglect again terms which scale as $\eta g_{\rm eff}$ or $\mu$. Eq.~(\ref{ij1}) describes relatively fast oscillations on the time scale given by the phonon frequency $\nu$. As we shall see in the next subsection of this appendix, on the relatively slow time scale we can therefore approximate these coherences, which depend on the slowly evolving variable $\zeta$, by their average values.

To obtain a closed set of effective cooling equations, we also need to have a closer look at the $y$ and the $\tilde y$ expectation values in Eq.~(\ref{mbetadot}). Doing so, we find, for example, that 
\begin{eqnarray} \label{split3} 
\dot y_{222} &=& - 3 \nu \, y_{232} \, ,
\end{eqnarray}
while
\begin{eqnarray} \label{split4} 
\dot {\tilde y}_{222} &=& - \nu \left( \tilde y_{223} + 2 \tilde y_{232} \right) \, , \nonumber \\
\dot {\tilde y}_{233} &=& \nu \left( \tilde y_{223} + \tilde y_{232} - \tilde y_{333} \right) \, , \nonumber \\
\dot {\tilde y}_{332} &=& \nu \left( 2 \tilde y_{232} - \tilde y_{333} \right) \, .  
\end{eqnarray} 
These expectation values all oscillate essentially rapidly around zero.

\subsection{Adiabatic elimination} \label{xcoherences2}

To finally obtain an effective cooling equation for the time evolution of the slowly evolving phonon expectation values $m$ and $\zeta$, we now determine the $x$, $y$, $\tilde y$ and the $\tilde z$-expectation values in Eq.~(\ref{mbetadot}) via an adiabatic elimination. Considering the parameter regime in Eq.~(\ref{condi2})  and setting the time derivatives in Eq.~(\ref{ij1}) equal to zero, we find for example that
\begin{eqnarray} \label{C8}
\tilde z_{23} = \tilde z_{32} &=&  0 \, , \nonumber \\
\tilde z_{22} &=& \tilde z_{33} \, .
\end{eqnarray}
Comparing $\zeta$ in Eq.~(\ref{notationfinal}) with the definition of $\tilde z_{ab}$ in Eq.~(\ref{notation500}), we moreover find that $\zeta = (\tilde z_{22} + \tilde z_{33})/4$ which implies
\begin{eqnarray} \label{C9}
\tilde z_{22} = \tilde z_{33} &=& 2 \zeta \, .
\end{eqnarray}
Taking this into account when adiabatically eliminating the $x_{ab}$ variables by setting the time derivatives in Eq.~(\ref{ij0}) equal to zero yields
\begin{eqnarray} \label{C10}
x_{22} &=&  - {8N \eta g_{\rm eff} \kappa \left[ \kappa^2 + 4 \left( \delta^2 + \nu^2 \right) \right] \over 
  \kappa^4 + 16 \left( \delta^2 - \nu^2 \right)^2 + 8 \kappa^2 \left( \delta^2 + \nu^2 \right)} \, \zeta \, , \nonumber \\
x_{23} &=& - {16 N \eta g_{\rm eff} \delta \left[ \kappa^2 + 4 \left( \delta^2 - \nu^2 \right) \right] \over 
   \kappa^4 + 16 \left( \delta^2 - \nu^2 \right)^2 + 8 \kappa^2 \left( \delta^2 + \nu^2 \right)} \, \zeta \, , \nonumber \\
x_{32} &=& - {16 N \eta g_{\rm eff} \nu \left[ \kappa^2 -4 \left( \delta^2 - 4 \nu^2 \right) \right]\over
   \kappa^4 + 16 \left( \delta^2 - \nu^2 \right)^2 + 8 \kappa^2 \left( \delta^2 + \nu^2 \right)} \, \zeta \, , \nonumber \\
x_{33} &=& - {64 N \eta g_{\rm eff} \kappa \delta \nu \over
   \kappa^4 + 16 \left( \delta^2 - \nu^2 \right)^2 + 8 \kappa^2 \left( \delta^2 + \nu^2 \right)} \, \zeta \, . ~~~~~~
\end{eqnarray}
This result can be checked relatively easily by substituting Eqs.~(\ref{C9}) and (\ref{C10}) back into Eq.~(\ref{ij0}). Finally, setting the right hand sides of Eqs.~(\ref{split3}) and (\ref{split4}) equal to zero, yields
\begin{eqnarray} \label{50}
y_{232} = \tilde y_{223} &=& 0
\end{eqnarray}
up to terms which scale as $\mu/\nu$. Combining the results of this subsection with Eq.~(\ref{mbetadot}) eventually yields the effective cooling dynamics in Eq.~(\ref{erna}).


\begin{thebibliography}{100}
\bibitem{footnote}
The collective cavity cooling scheme which we propose here also operates in the bad-cavity limit where $g$, $\kappa$ and $\Gamma$ are all of similar size \cite{Kimold}. However making the assumption below simplifies the following calculations and helps us to point out its underlying mechanisms more clearly.
\end{thebibliography}
\end{document}